# PyRod – Tracing Water Molecules in Molecular Dynamics Simulations

*David Schaller[†], Szymon Pach[†] and Gerhard Wolber[†],\**

[†] Pharmaceutical and Medicinal Chemistry, Freie Universität Berlin, Königin-Luise-Str. 2+4, 14195 Berlin, Germany

ABSTRACT Ligands entering a protein binding pocket essentially compete with water molecules for binding to the protein. Hence, the location and thermodynamic properties of water molecules in protein structures have gained increased attention in the drug design community. Including corresponding data into 3D pharmacophore modeling is essential for efficient high throughput virtual screening. Here, we present PyRod, a free and open-source python software that allows for visualization of pharmacophoric binding pocket characteristics, identification of hot spots for ligand binding and subsequent generation of pharmacophore features for virtual screening. The implemented routines analyze the protein environment of water molecules in molecular dynamics (MD) simulations and can differentiate between hydrogen bonded waters as well as waters in a protein environment of hydrophobic, charged or aromatic atom groups. The gathered information is further processed to generate dynamic molecular interaction fields (dMIFs) for visualization and pharmacophoric features for virtual screening. The described software was applied to 5 therapeutically relevant drug targets and generated pharmacophores were evaluated using DUD-E benchmarking sets. The best performing pharmacophore was found for the HIV1 protease with an early enrichment factor of 54.6. PyRod adds a new perspective to structure-based screening campaigns by providing easy-to-interpret dMIFs and purely protein-based pharmacophores that are solely based on tracing water molecules in MD simulations. Since structural information about co-crystallized ligands is not needed, screening campaigns can be followed, for which less or no ligand information is available. PyRod is freely available at https://github.com/schallerdavid/pyrod.

INTRODUCTION

Unliganded protein binding pockets are occupied by water molecules which obligates potential ligands to compete with these water molecules for binding to the protein. Hydrogen bonds between water and protein need to be broken and replaced water molecules will be released to the bulk solvent. This process heavily affects the thermodynamic properties of the system and renders water as one of the key elements to understand and promote ligand binding[1,2]. Several approaches (e.g. 3D-RISM[3], GIST[4]) have been developed and employed to estimate the enthalpic and entropic contribution of replacing water molecules from protein binding sites which proofed to be useful in pinpointing hot spots for ligand binding and in explaining structure-activity relationships. Including data from molecular dynamics (MD) simulations was found to improve such predictions[5]. However, researchers at GSK conclude in a recent perspective[6] that many studies utilizing water-based methods are of retrospective nature and several results could have been obtained by simply looking at the atomistic models, e.g. growing a ligand into a hydrophobic protein pocket will most likely increase the affinity.

Pharmacophores describe electrostatic and steric features needed for a molecule to bind to a desired drug target and can be employed in a truly prospective fashion in efficient high-throughput virtual screening campaigns to identify novel active entities[7]. Recently, MD simulations were analyzed to generate so called water pharmacophores[8]. The researchers analyzed the thermodynamic characteristics of hydration sites in binding pockets of several drug targets and were able to translate this information into pharmacophores that were successfully evaluated in retrospective screening campaigns. However, the method makes use of commercial software and is not available for public use.

Here, we present PyRod, a free and open-source python software that was built to translate the highly complex, but important information from MD simulations into simplistic and highly efficient pharmacophore models suitable for virtual screening. PyRod supports the 4 major forcefields CHARMM[9], AMBER[10], GROMOS[11] and OPLS[12] granting maximum flexibility for the user in choosing the simulation package for generating MD simulation data. We applied PyRod to 5 important drug targets and evaluated its capability to generate successful pharmacophores for virtual screening.

IMPLEMENTATION

PyRod is available as open-source software written in python 3[13] and employs the external packages MDAnalysis[14], NumPy[15] and SciPy[16]. It is composed of several components that can be executed individually via self-explainable config files (Figure 1A). Additionally, a trajectory pharmacophore combo config file is provided which enables a one-step-execution of several tasks.

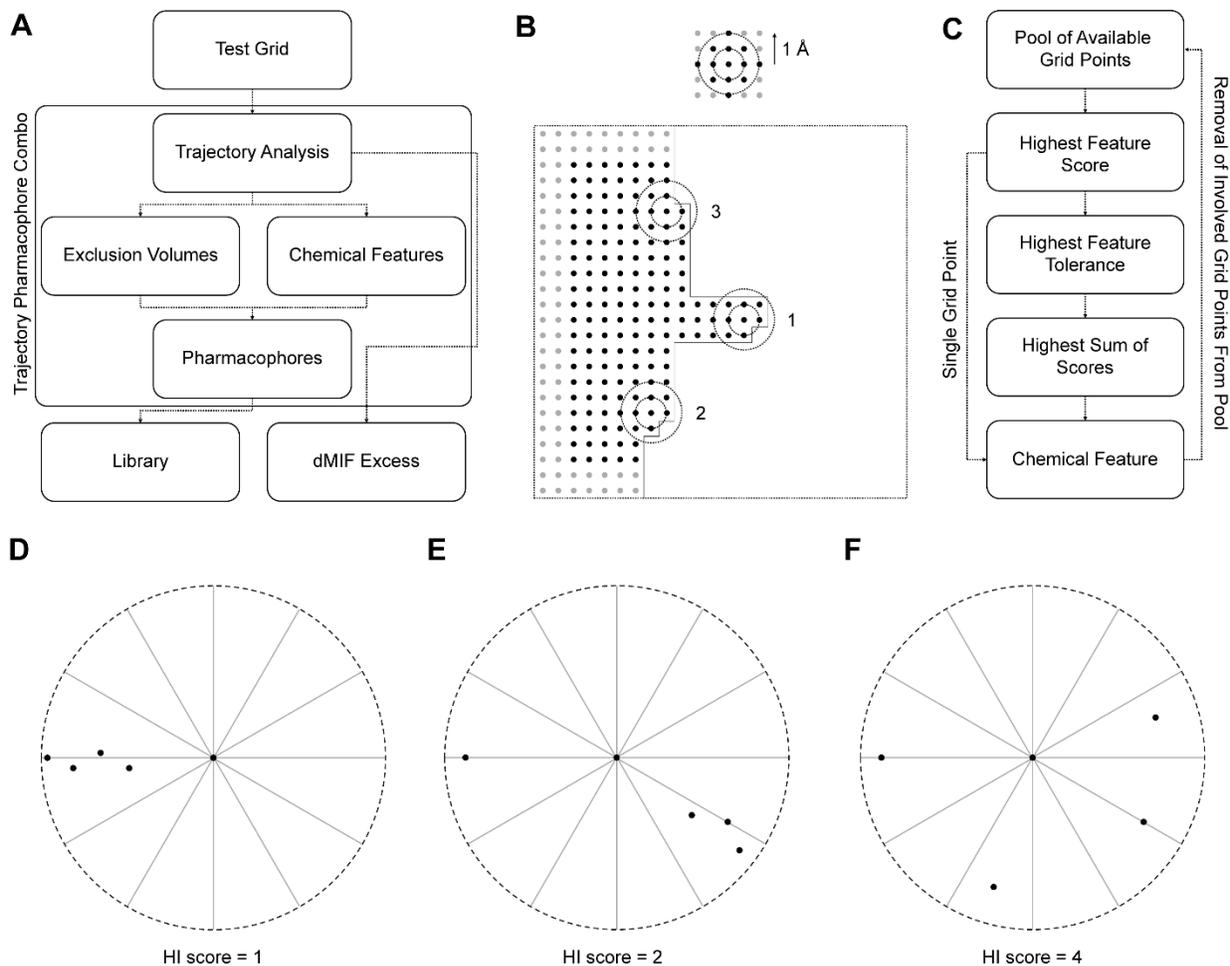

**Figure 1:** (A) Workflow diagram of PyRod. Each box represents a single component of PyRod that can be executed separately or in combination using the trajectory pharmacophore combo. (B) Depiction of the exclusion volume sphere generation algorithm. Top: The neighborhood within 0.5 Å and 1 Å of each grid point is analyzed. Bottom: Only grid points with a shape score less than 1 are depicted, representing an area with low water occupancy, e.g. within the protein. Grid points with a shape score of 1 or higher are not depicted and would fall into the white area corresponding to a potential protein binding pocket. Grey grid points are at the grid boundaries and are not considered for exclusion volume placement. The described algorithm favors exclusion volume spheres with a maximum number of neighbors within 0.5 Å, but a minimal number of neighbors within 1 Å, and would generate 3 exclusion volumes in this example in the order described by the numbers. (C) Depiction of the iterative chemical feature generation. (D-F) Examples for hydrophobic interaction (HI) scores scaled by buriedness. Each depiction represents a water oxygen with 4 hydrophobic atoms within 5 Å resulting in different HI scores. The buriedness algorithm evaluates the hydrophobic atom positions by calculating angles with the water oxygen as vertex. (D) Water molecule is not buried. (E) Water molecule is buried between two hydrophobic centers. (F) Water molecule is buried between 4 hydrophobic centers.

**Test Grid.** This component facilitates the identification of parameters for proper grid placement allowing the user to focus on the protein area of interest in later trajectory analysis. The x, y and z center parameters are used to define the center of the grid which can be retrieved from e.g. a central atom in the binding pocket with coordinates stored in the topology file or by employing pocket detection algorithms externally. The x, y and z edge length parameters define the size of the grid and are usually set between 20 and 30 Å, but can be set higher if the whole protein surface should be explored. The test grid will be saved in pdb format with grid points as pseudo atoms. This file can then be visualized together with the topology file to improve parameters if required.

**Trajectory Analysis.** The implemented routines analyze the protein environment of water molecules in MD simulations to predict favorable sites for chemical feature placement, i.e. hydrogen bond, ionizable, hydrophobic and aromatic interactions. The employed heuristic scoring functions do not calculate thermodynamic properties, but instead estimate favorable regions for chemical feature placement in each frame based on fast-to-calculate geometric descriptors. Since scoring is only performed in presence of water, PyRod favors regions with stable water molecules whose replacement by ligands will result in a gain of entropy. Chemical feature scores are summated throughout the trajectory in a NumPy[15] representation of the 3D grid whose position and size can be determined by using the test grid component. Spacing between grid points is fixed to 0.5 Å. MDAnalysis[14] is employed as topology and trajectory reader supporting various molecular data formats. Residue and atom names are standardized enabling the support of widely used force fields CHARMM[9], AMBER[10], GROMOS[11] and OPLS[12]. The gathered information is transformed for each chemical feature type into dynamic molecular interaction fields (dMIFs) and can be saved in density map format (kont, cns, xplor) for visualization of pharmacophoric binding pocket characteristics. The name dMIF was chosen, since the generated maps introduce dynamics to the concept of molecular interaction fields (MIFs), an established tool in modern drug discovery describing the interaction energy between a probe and a molecular target. For an extensive review on molecular interaction fields we would like to refer the reader's attention to a publication by Artese and co-authors[17].

In a first step, the protein is analyzed for atoms and atom groups corresponding to potential chemical feature interaction partners (supporting information Tab. S1), e.g. oxygen atoms of the aspartate carboxylate group are hydrogen bond acceptors and part of a negatively charged group if deprotonated. Next, water molecules are localized in each trajectory frame and their protein environment is analyzed for the previously defined chemical feature interaction partners. If certain geometrical criteria are met (supporting information Tab. S2), grid points within 1.41 Å (radius of water molecule[18]) of the water molecule are identified using fast KDTree[19] routines from SciPy[16] and scored according to the chemical feature type.

Water molecules have two lone pairs and two hydrogens allowing the formation of hydrogen bond as acceptor as well as donor. If water molecules at a certain position are half of the time hydrogen bonded to the protein as donor and half of the time as donor and acceptor, this could indicate two different protein conformations which is important to pharmacophore generation.

Hence, hydrogen bonding interactions are split into 6 categories, i.e. single hydrogen bond donor (HD), single hydrogen bond acceptor (HA), double hydrogen bond donor (HD2), double hydrogen bond acceptor (HA2) and mixed hydrogen bond donor/acceptor (HDA). Water molecules involved in more than two hydrogen bonds with the protein are treated as trapped water (TW). Such water molecules are typically deeply buried in the protein making them barely accessible for ligands and only very few ligands would be able to fulfill the geometric criteria to replace the water molecule sufficiently. Thus, trapped waters are not considered for later chemical feature generation. However, trapped waters can be of interest in later screening hit selection, since they might serve as bridge between protein and ligand. Water molecules near metal ions (e.g. $Zn^{2+}$, $Mn^{2+}$) are treated as hydrogen bond acceptors and are included in the hydrogen bond count to identify trapped waters. Positions of the protein interaction partners are stored to allow later chemical feature generation with directionality. Gathered hydrogen bond scores are transformed into easy-to-interpret occupancies, e.g. a HA score of 15 means that in 15 % of the frames there was a water molecule at this position being involved as a single hydrogen bond acceptor with the protein.

Positive (PI) and negative ionizable interactions (NI) are also scored as occupancies but are additionally scaled by distance, since they represent long range interactions whose energy decays with increasing distance. Furthermore, PI and NI quench each other. A PI score of 80 can describe very different situations, e.g. there was a water molecule in 80 % of the frames in the optimal distance to 1 PI partner but no NI partner or in 40 % of the frames in the optimal distance to 3 PI partners and 1 NI partner.

Aromatic interactions (AI) show a rather complex geometry involving several angles and distances (supporting information Tab. S2). Additionally, in contrast to hydrogen bond and ionizable interactions, water molecules will not necessarily accumulate at a favorable position for potential ligand interaction partners. Thus, grid points close to such water molecule are evaluated individually to satisfy the AI geometry criteria and receive an individually distance scaled occupancy score. Idealized positions for potential interaction partners are stored for later feature generation with directionality.

Cation-π interactions are also recorded by the implemented routines and included in PI and AI scores, e.g. water molecules close to the aromatic ring of a phenylalanine will be scored for PI and water molecules close to the positively charged amine of a lysine will be scored for AI. However, they receive a dedicated heuristic scoring function differing from earlier presented PI and AI scoring (supporting information Tab. S2).

Regions for potential hydrophobic interactions (HI) are identified by counting hydrophobic atoms in vicinity of each water molecule. This crude atom count is additionally scaled by buriedness to highlight regions with water molecules deeply enclosed in a hydrophobic pocket (Figure 1D-F). In a first step, positions of hydrophobic atoms within 5 Å of the water oxygen are collected. If only one hydrophobic atom was identified, the hydrophobic score is 1 and no further calculation is performed. Otherwise, hydrophobic atom positions are analyzed to estimate the buriedness of the water molecule as follows. Two hydrophobic atom positions are determined that form the maximal angle with the water oxygen position as vertex, e.g. two hydrophobic atoms with a water molecule exactly in between would lead to an angle of 180 degrees. If this maximal angle is less than 30 degrees, the algorithm is terminated, and the hydrophobic score remains 1 (Figure 1D). Such situation corresponds to a geometry, where a water molecule is close to multiple hydrophobic atoms but not buried. Otherwise, the hydrophobic score is increased by 1, both hydrophobic atom positions are marked as accepted and one of the two hydrophobic atom positions forming the maximal angle is randomly selected as reference position for further processing. The remaining hydrophobic atom positions are analyzed in an iterative fashion as follows. First, the hydrophobic atom position is identified that forms the maximal angle with the reference position and the water oxygen position as vertex. Next, this hydrophobic atom position is evaluated for angles formed with the already accepted positions and the water oxygen as vertex. If none of the formed angle is smaller than 30 degrees, the hydrophobic score is increased by 1 and the evaluated position is marked as accepted position. Otherwise, this position is ignored. This procedure is repeated until all hydrophobic atom positions were evaluated resulting in a hydrophobic score that is strongly dependent on the hydrophobic buriedness of the water molecule. However, highly hydrophobic regions may not be sampled well by water molecules and will be scored less frequently and consequently may receive a lower absolute HI score. Hence, a normalized score is provided as well ($HI_{norm}$) which reports the average hydrophobic score per occurrence, e.g. a $HI_{norm}$ score of 5.3 means that when a water molecule occurred at this position the near grid points retrieved on average a HI score of 5.3.

**Exclusion Volume Spheres.** During trajectory analysis the presence of water is recorded to generate a shape dMIF. It is used in this component to place exclusion volumes limiting the binding pocket volume in later pharmacophores with the following algorithm (Figure 1B). Grid points must have a shape score less than 1 (corresponding to water occupancy of 1 %, can be changed by the user) and are sorted for the number of neighbors (grid points with a shape score less than 1) within 1 Å. Next, each grid point is evaluated as center of an exclusion volume starting with the grid point with the lowest number of neighbors within 1 Å. Grid points with very few neighbors usually correspond to protein side chains pointing inside the binding pocket and are prioritized by the algorithm. To be accepted as the center of an exclusion volume, a grid point must not be at the grid boundaries, must have exactly 7 neighbors within 0.5 Å but less than 33 neighbors within 1 Å ensuring that exclusion volumes are placed only at the interface of protein and water but not too close to the chemical features and finally, must not be within 4 Å (2 Å if restrictive mode is enabled) of an already generated exclusion volume.

**Chemical Features.** A novel algorithm was implemented translating dMIFs into corresponding chemical features for pharmacophore virtual screening (Figure 1C). First, all grid points become part of a pool of available grid points for the respective chemical feature generation. Next, the grid points with the highest feature score in the grid point pool are determined. If this search results in a single grid point, its position will be used as center of the chemical feature. The tolerance radius of that chemical feature is identified by iteratively increasing the search radius (minimum=1.5 Å) from the feature center in 0.5 Å steps. If the feature score of a grid point within the search radius is below half of the current highest feature score, the search is stopped, and the current search radius will be used as tolerance radius for the chemical feature. If multiple grid points share the highest feature score in the grid point pool, the following procedures are performed to select a single grid point as feature center. Tolerance radii are calculated for each of the considered grid points. The grid point with the highest tolerance radius will be used as center of the chemical feature. If multiple grid points share the highest tolerance radius, the sum of feature scores of the grid points within the tolerance radius is calculated and the grid point with the highest sum of feature scores is selected as feature center. If this procedure does not lead to the selection of a single grid point, a random pick of the remaining grid points is performed. Grid points within the tolerance radius of a chemical feature must not be part of an already generated chemical feature of the respective chemical feature type.

This criterion prevents overlap of chemical features within chemical feature types. In case of hydrogen bond and aromatic interactions, recorded positions of interaction partners are clustered by searching for the interaction partner position with the most neighboring interaction partner positions within 1.5 Å. This procedure allows the generation of chemical features with directionality. Finally, grid points within the tolerance radius of the feature center are removed from the pool of available grid points and a new iteration is started. The chemical feature generation is terminated if 20 chemical features (can be changed by the user) of the respective chemical feature type were generated or if the highest feature score of the grid point pool decreases below 1.

**Pharmacophores.** The output of exclusion volume and chemical feature generation is merged and saved as a single "super pharmacophore" containing all previously generated chemical features and exclusion volumes. Additionally, a pharmacophore can be saved containing only the highest ranked features for each chemical feature class as specified by the user, e.g. 20 highest scored hydrogen bonding features and 5 highest scored hydrophobic features. Currently, PyRod supports LigandScout[20] and pdb-like pharmacophore formats. The pdb-like pharmacophore file uses the residue name column to specify the chemical feature type and aims at providing a pharmacophore format readable by human and various molecular modeling softwares. However, directionality is not included in the pdb-like pharmacophore format.

**Combinatorial Library.** The generated "super pharmacophores" can contain more than 100 chemical features, which remains computationally challenging to screen also with the current progress in CPU performance. Thus, reducing the number of chemical features is key to enable fast high-throughput virtual screening. This component facilitates the generation of a combinatorial library of pharmacophores with a specified number of chemical features as defined by the user. First, the user preselects chemical features of interest in LigandScout[20] and saves this pharmacophore for combinatorial processing. Chemical features that should be present in every generated pharmacophore have to be set mandatory, whereas chemical features that should be added in a combinatorial fashion have to be set optional. Next, the user can specify the limits for minimal and maximal number of chemical features in the config file, i.e. number of independent chemical features, number of hydrogen bonding features, number of ionizable features, number of aromatic features and number of hydrophobic features. Prior to library generation the user will be informed about the number of possible pharmacophores and prompted for execution. To further limit the library size each pharmacophore is evaluated for the following rules, i.e. (i) ionizable and hydrophobic features should not appear within 3 Å, (ii) different hydrogen bonding features should not be present within 1.5 Å, since such situation implies two different protein conformations, (iii) different ionizable features should not be present within 3 Å and (iv) hydrogen bonding features of HA2, HD2 and HDA are not allowed to be split. PyRod also provides a customizable pharmacophore evaluation script written in python performing receiver operatic characteristics analysis with LigandScout[20].

**dMIF Excess.** Selectivity within a protein family as well as the occurrence of mutation-induced resistance remain a major challenge in modern drug discovery[21,22]. It would be desirable to exploit such minor differences in protein binding pockets. Thus, this component enables the comparison of dMIFs between closely related proteins by generating dMIF excess maps visualizing the excess of one system over the other.

TEST SYSTEMS

PyRod performance was evaluated on 5 important drug target test systems, i.e. cyclin-dependent kinase 2-cyclin A complex (CDK2, 5if1[23] (1)), HIV-1 protease (HIV1P, 1nh0[24] (2)), estrogen receptor alpha (ERα, 1xpc[25] (3)), dopamine D3 receptor (D3R, 3pbl[26] (4)) and adenosine A$_{2A}$ receptor (A$_{2A}$R, 5iu4[27] (5)). Protein selection was based on therapeutic relevance, availability of benchmarking sets from DUD-E[28] and crystal structures from PDB[29] as well as protein family diversity.

**System Setup.** Crystal structures were retrieved from PDB[29] and prepared in MOE 2015[30] as follows. Ligands were deleted as well as water more than 5 A away from the protein. Errors were corrected with the Structure Preparation tool. The low resolution D3R structure 3pbl misses a sodium ion that is known to be crucial for inactive class A GPCR states[31]. Hence the sodium ion and 6 coordinating water molecules were transferred from the high resolution structure of δ opioid receptor (4n6h[32]) into the D3R system. Chain breaks were capped with ACE and NME. Protonation states were assigned using Protonate 3D tool at pH 7. Non-membrane proteins (CDK2, HIV1P and ERα) were solvated in a cubic box with TIP4P water, 0.15 M NaCl and 10 Å padding using Maestro 11.3[33]. Membrane proteins (A$_{2A}$R and D3R) were embedded in a POPC bilayer according to the orientation provided by the PPM server[34] and solvated in an orthorhombic box with TIP4P water, 0.15 M NaCl and 10 Å padding.

**Molecular dynamics simulation.** Simulations were performed with Desmond 5.1[35] and the OPLS 2005[36] forcefield on a Nvidia GeForce GTX 1070 graphics card. Minimization and equilibration were performed with default settings. 10 replications of 10 ns simulation were performed for each system with periodic boundary conditions in NPT ensemble. The temperature was retained at 300 K using the Nose-Hoover thermostat, and the pressure at 1.01325 bar using the Martyna-Tobias-Klein barostat. Coordinates were saved every 5 ps resulting in 2000 frames per simulation. Trajectories were additionally processed in VMD 1.9.2[37], i.e. the protein was centered in the water box by using the pbc tool and trajectories were aligned on the protein backbone heavy atoms using the RMSD Trajectory tool.

The CDK2 system (5if1[23]) was also simulated with OpenMM 7.2.2[38] on a Nvidia GeForce GTX 1070 graphics card employing the Amber forcefield ff14SB[10] with TIP4P-Ew water model to test the effect of restraining heavy atoms. The same prepared protein structure was used as for Desmond simulations described above. The protein was solvated in a cubic water box with 10 Å padding and 0.15 M NaCl. The Particle Mesh Ewald method was used to calculate long range electrostatic interactions with a 10 Å cutoff and all bonds involving hydrogens were constrained in length. Langevin dynamics were performed at 300 K with 2 fs time step. 10 replications of 10 ns simulations were performed with periodic boundary conditions and NPT ensemble. Coordinates were saved every 5 ps resulting in 2000 frames per simulation. The simulations were performed with and without a custom force of 5 kcal restraining protein heavy atoms at their initial position. Resulting trajectories were processed in VMD[37] as already described.

**PyRod.** Grid parameters were adjusted using the test grid component to center the grid in the binding site. Grids were cubic with edge lengths of 20 Å for ERα and CDK2, 25 Å for D3R and HIV1P and 30 Å for $A_{2A}R$. Trajectories were processed with PyRod 0.7.1 using last 5 ns of each replication resulting in 10,000 frames for analysis of each system with default settings. Generated dMIFs were visualized and analyzed in LigandScout 4.2[20] to preselect pharmacophore features according to feature scores and their arrangement in the binding pocket. Selected chemical features were subjected to combinatorial processing with the combinatorial library component of PyRod. Chemical feature limits for each target can be found in the supporting information (Tab. S3).

**Pharmacophore screening.** The ligand benchmarking sets with actives and decoys for CDK2, HIV1P, ERα, D3R and $A_{2A}R$ were retrieved from DUD-E server[28] in SMILES format. For CDK2, ERα, D3R and $A_{2A}R$ 25 conformations were generated per molecule with iCon as implemented in LigandScout 4.2[20]. For HIV1P 200 conformations were generated per molecule, since the active set primarily contains peptidomimetics with many rotatable bonds. These databases were used for pharmacophore evaluation in LigandScout 4.2[20] employing receiver operating characteristic (ROC) curves.

RESULTS & DISCUSSION

**CDK2.** The ATP binding pocket of CDK2 is a well characterized site for inhibition with a plethora of crystal structures deposited in the PDB. The most frequently observed interactions include hydrogen bonds formed with the backbone of residues E81 and L83[39]. Concordantly, PyRod identified a hydration site at which water molecules act as single hydrogen bond donor to the backbone oxygen of E81 in 63 % of all frames (Figure 2B). Besides being involved in a single hydrogen bond, these water molecules are also in a very hydrophobic environment (HI score=420, $HI_{norm}$ score=5.85). These characteristics render this hydration site as an essential position for ligand binding, since replacing restrained water molecules from a hydrophobic pocket by a corresponding ligand moiety should be beneficial for the entropy and enthalpy of the system. Adjacent to E81 are further hydration sites at which water molecules are involved in hydrogen bonds with the backbone of L83 in 40 % of the frames as single donor, in 19 % of the frames as single acceptor and in 15 % of the frames as mixed donor and acceptor (Figure 2B). Likewise, these hydration sites lie in a hydrophobic environment (HI score=180-350, $HI_{norm}$ score=3.5-6.25) highlighting these positions for additional ligand interactions. Several hotspots were identified for placing positive ionizable groups, i.e. at the interface of D145 and F80 as well as next to D86 with PI scores of 70 and 25 respectively, and aromatic moieties, i.e. close to the salt bridge formed by K33 and D145 as well as adjacent to F80 with AI scores of 45 and 70 respectively (Figure 2C). A hydrophobic band with HI scores ranging from 100 to 300 is spanning the binding pocket that resulted in the generation of 6 hydrophobic features (Figure 2C). In total 15 chemical features were selected based on the corresponding feature score and their arrangement in the binding pocket (Figure 2D). By employing the combinatorial library component of PyRod, these features were combined to 816 pharmacophores with 3 to 5 chemical features. The hydrogen bonding donor feature pointing towards the backbone oxygen of E81 was selected to be present in every pharmacophore. Further parameters can be found in the supporting information (Tab. S3). Using LigandScout[20] all pharmacophores were screened against an active set retrieved from the DUD-E[28] database and additionally evaluated against decoys if 5 % of the actives were found. Hit lists were evaluated for early enrichment factor (EF1) and plotted against found actives to select pharmacophores of interest (Figure 2E). The most selective pharmacophore (EF1=30.3) consists of 2 hydrophobic features, 1 hydrogen bond donor and 1 hydrogen bond mixed donor/acceptor (Figure 2F).

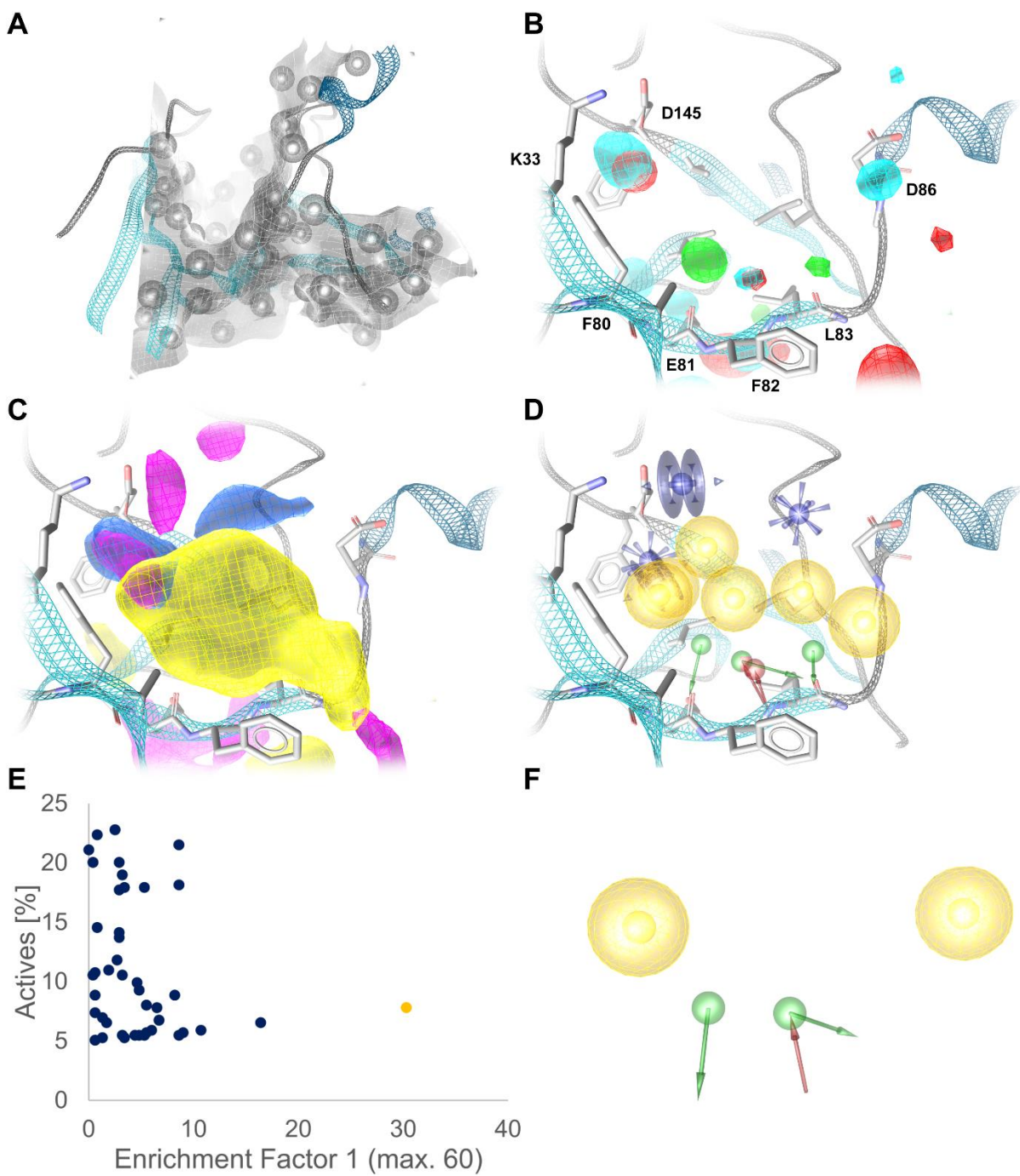

**Figure 2:** (A) Shape dMIF of CDK2 (cutoff 1) with exclusion volumes present in each generated pharmacophore. (B, C) Characterized binding pocket with dMIFs for single hydrogen bond donor (green, cutoff 38), single hydrogen bond acceptor (red, cutoff 17), mixed hydrogen bond donor/acceptor (cyan, cutoff 14), positive ionizable (blue, cutoff 27), aromatic interaction

(magenta, cutoff 36) and hydrophobic interaction (yellow, cutoff 100). Cutoffs were chosen to visualize decision making in subsequent pharmacophore feature selection. (D) Selected pharmacophore features based on chemical feature score and arrangement (green arrow – hydrogen bond donor, red arrow – hydrogen bond acceptor, yellow sphere – hydrophobic interaction, blue star – positive ionizable, blue ring plane – aromatic interaction). (E) Performance evaluation of pharmacophore library. (F) Pharmacophore with best early enrichment factor ($EF_{1;5;10;100}$: 30.3;30.3;30.3;30.3, $AUC_{1;5;10;100}$: 0.99;0.99;0.88;0.54).

**D3R.** Currently, only a single crystal structure of D3R is available. However, this target was studied extensively with many known ligands stored in public data bases. A key interaction shared across all aminergic GPCRs is a charged interaction to an aspartate in the orthosteric binding pocket[40]. PyRod located several hydration sites with water molecules pointing as single hydrogen bond donor towards D110 in 35-50 % of the frames with PI scores between 35 and 50 (Figure 3B-C). Additionally, we found 2 sites with water molecules acting as hydrogen bond acceptor with a HA score of 20 next to the backbone nitrogen of I183 and to the imidazole ring of H349. Hydrophobic hotspots (HI score=140-225, $HI_{norm}$ score=3.80-4.25) were identified close to F345 and above D110 as well as sites for aromatic interactions (AI score=15) next to F345. In total, 16 chemical features were selected and combined to 2441 pharmacophores with 3 to 5 independent features (Figure 3D). Further parameters can be found in the supporting information (Tab. S3). The best performing pharmacophore (EF1=7.7) consists of 1 hydrophobic feature, 1 hydrogen bond acceptor and 1 positive ionizable feature (Figure 3E-F).

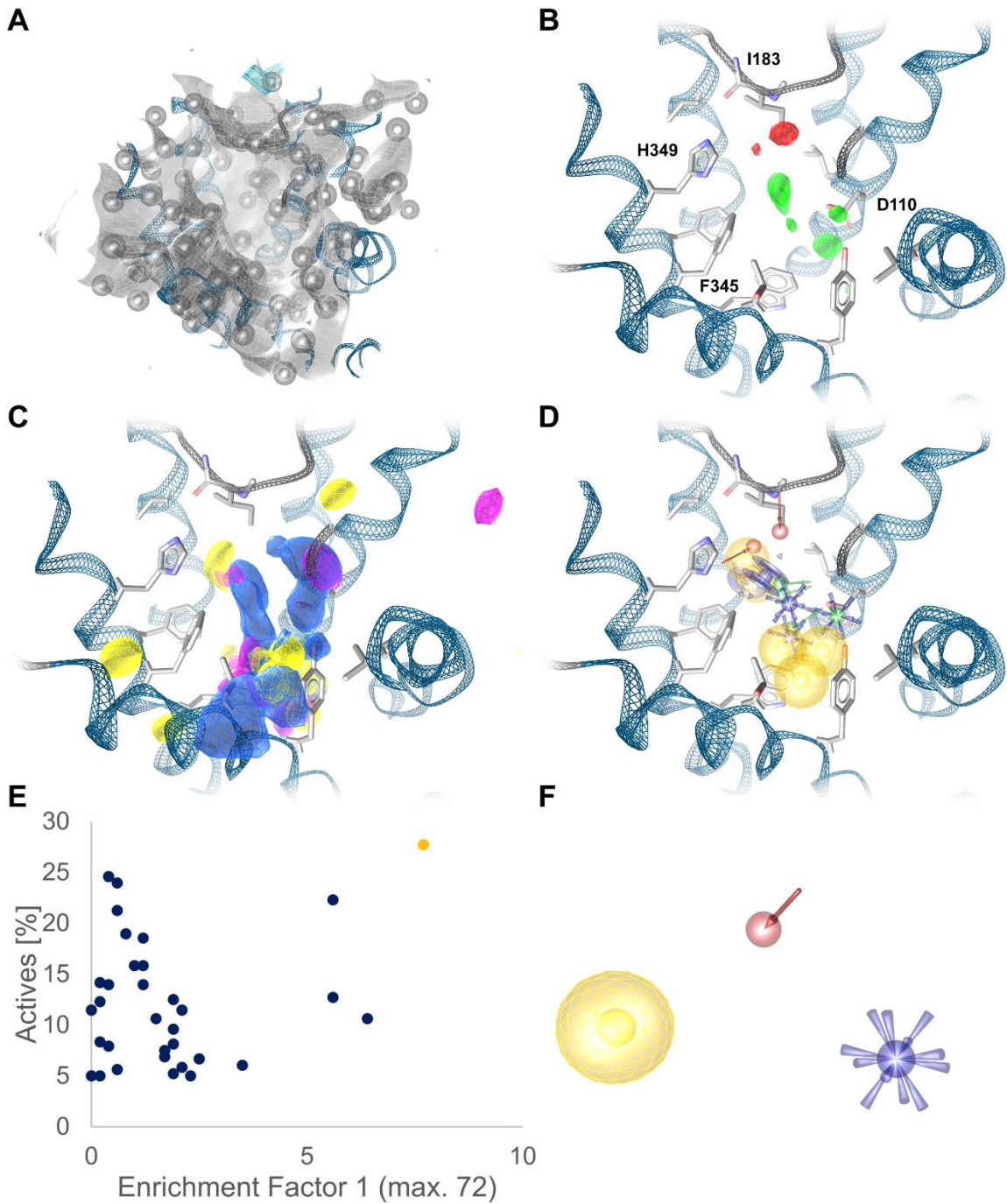

**Figure 3:** (A) Shape dMIF of D3R (cutoff 1) with exclusion volumes present in each generated pharmacophore. (B, C) Characterized binding pocket with dMIFs for single hydrogen bond donor (green, cutoff 36), single hydrogen bond acceptor (red, cutoff 19), positive ionizable (blue, cutoff 33), aromatic interaction (magenta, cutoff 14) and hydrophobic interaction (yellow, cutoff

160). Cutoffs were chosen to visualize decision making in subsequent pharmacophore feature selection. (D) Selected pharmacophore features based on chemical feature score and arrangement (green arrow – hydrogen bond donor, red arrow – hydrogen bond acceptor, yellow sphere – hydrophobic interaction, blue star – positive ionizable, blue ring plane – aromatic interaction). (E) Performance evaluation of pharmacophore library. (F) Pharmacophore with best early enrichment factor ($EF_{1;5;10;100}$: 7.7;3.3;2.3;2.0, $AUC_{1;5;10;100}$: 0.92;0.94;0.93;0.58).

**HIV1P.** The protease of HIV1 is a well characterized target for inhibiting virus replication. Mature HIV1P exists as a homodimer with two aspartates (D25, D25') in the catalytic center. The hydroxyl group of approved drugs mimics a water molecule present in the transition state and results in inhibition of the protease[41]. PyRod located two hydration sites between the two catalytic aspartates with HD2 scores of 40 and a PI score of 110 (Figure 4B). Hydrophobic pockets are symmetrically distributed around the catalytic center resulting in several hydrophobic features (HI score=120-270, $HI_{norm}$ score=2.80-4.90). PyRod also found the hydration site (Figure 4C) between the backbone nitrogens of I50 and I50' where water molecules are bound as double or single hydrogen bond acceptor (HA2 score=5, HA score=17). This water molecule is observed frequently in inhibitor-bound crystal structures serving as bridge between ligand and protein, but can also be replaced[41]. Finally, PyRod identified a hydration site next to D29 and D30 with water molecules acting as double hydrogen bond acceptor (HA2 score=6), single hydrogen bond acceptor (HA score=15) or mixed hydrogen bond donor/acceptor (HDA score=27). A single hydrogen bond acceptor feature was chosen at this position to represent all 3 observed water conformations (Figure 4D). 588 pharmacophores were generated by combining 15 pharmacophore features. The double hydrogen bond donor feature was selected to be present in every pharmacophore. Further parameters can be found in the supporting information (Tab. S3). The best performing pharmacophore ($EF_{1\%}$=54.6) consists of 3 hydrophobic features 2 hydrogen bond donors and 2 hydrogen bond acceptors (Figure 4E-F).

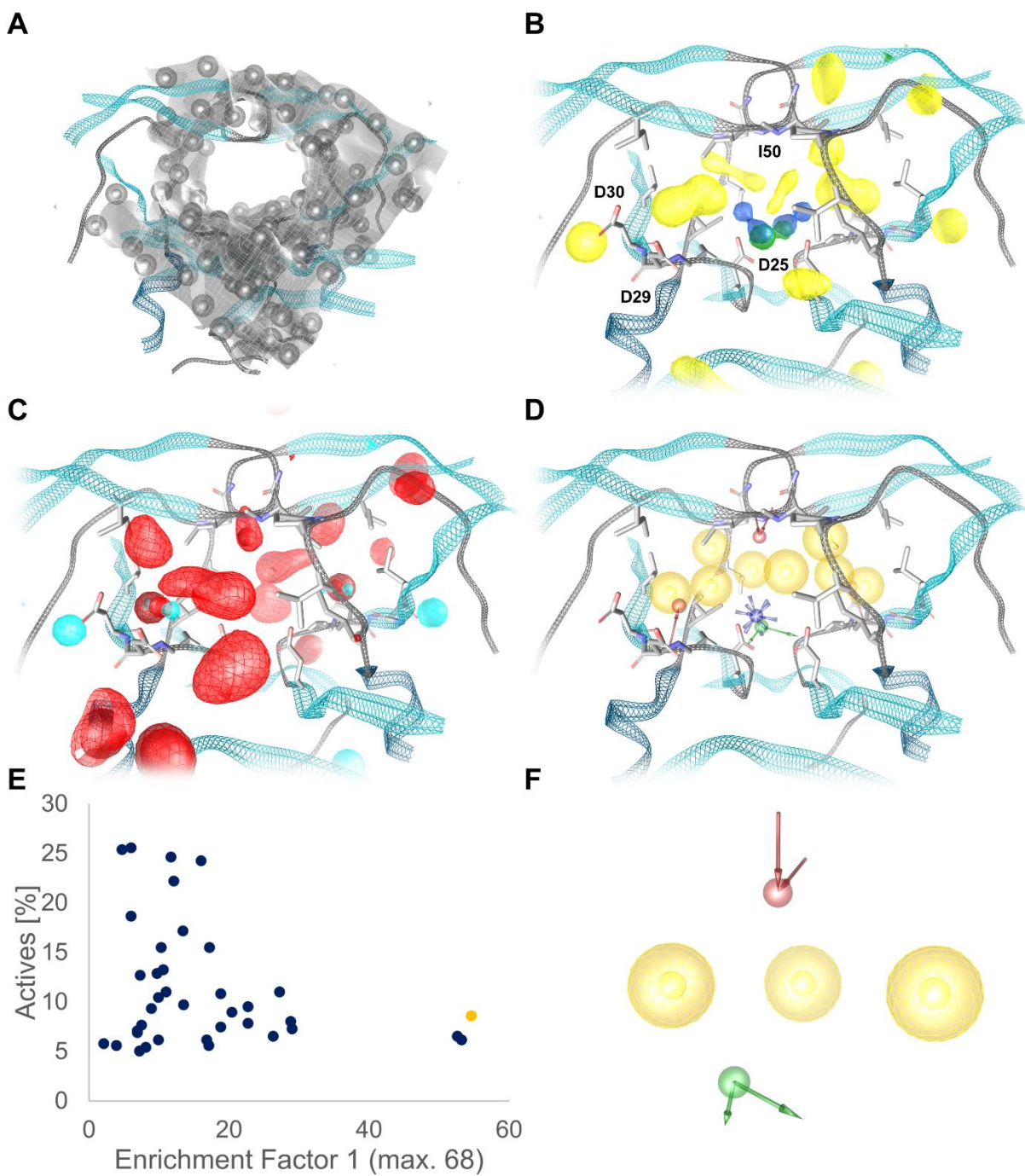

**Figure 4:** (A) Shape dMIF of HIV1P (cutoff 1) with exclusion volumes present in each generated pharmacophore. (B, C) Characterized binding pocket with dMIFs for double hydrogen

bond donor (dark green, cutoff 35), positive ionizable (blue, cutoff 90), hydrophobic interaction (yellow, cutoff 110), single hydrogen bond acceptor (red, cutoff 13), double hydrogen bond acceptor (dark red, cutoff 4) and mixed hydrogen bond donor/acceptor (cyan, cutoff 26). Cutoffs were chosen to visualize decision making in subsequent pharmacophore feature selection. (D) Selected pharmacophore features based on chemical feature score and arrangement (green arrow – hydrogen bond donor, red arrow – hydrogen bond acceptor, yellow sphere – hydrophobic interaction, blue star – positive ionizable). (E) Performance evaluation of pharmacophore library. (F) Pharmacophore with best early enrichment factor ($EF_{1;5;10;100}$: 54.6; 54.6; 54.6; 54.6, $AUC_{1;5;10;100}$: 1.00;1.00;0.93;0.54).

**ERα and $A_{2A}R$.** Estrogen receptor alpha and adenosine $A_{2A}$ receptor represent 2 test cases for which pharmacophore generation based on water dynamics was not successful. ERα contains a hydrophobic pocket[25] that is collapsing upon unrestrained molecular dynamics simulation. This ultimately leads to the placement of exclusion volumes at a position where co-crystallized ligands bind (Figure 5A). Agonists and antagonists of $A_{2A}R$ share two key interactions with F168 and N253[42]. The aromatic interaction with F168 is completely absent in the AI dMIF, since F168 is very flexible in the unbound state and adapts a conformation differing from the ligand bound state (Figure 5B). Also, N253 and E169 leave the ligand bound conformation quickly upon initiating unrestrained simulations and do not frequently interact with water molecules at positions known from ligand interaction with N253.

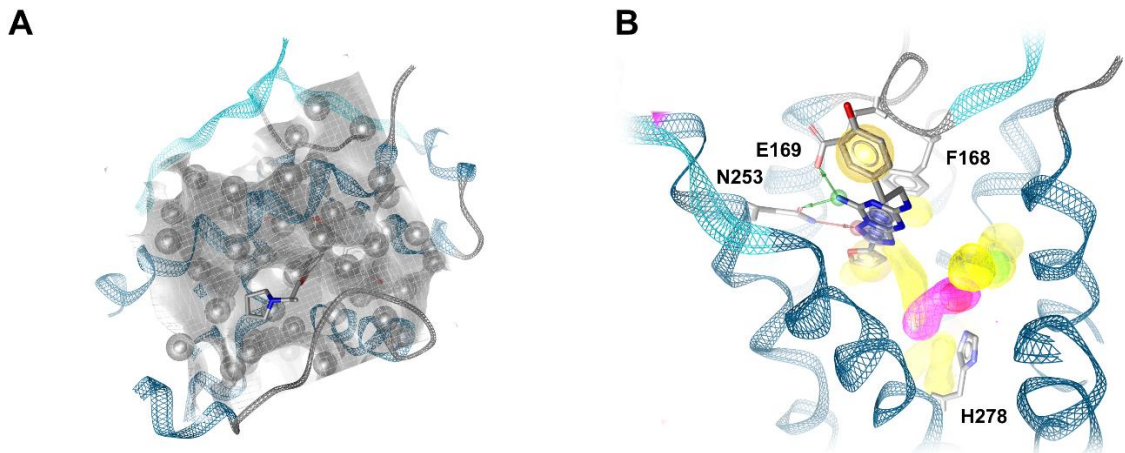

**Figure 5:** (A) Shape dMIF of ERα (cutoff 1) with an exclusion volume placed at a position where co-crystallized ligands bind. (B) Characterized binding pocket of $A_2AR$ with dMIFs for hydrogen bond donor (green, cutoff 40), single hydrogen bond acceptor (red, cutoff 40), aromatic interaction (magenta, cutoff 32) and hydrophobic interaction (yellow, cutoff 160). Essential key interactions known from co-crystallized ligands are not represented in the dMIFs (green arrow – hydrogen bond donor, red arrow – hydrogen bond acceptor, blue ring plane – aromatic interaction, yellow sphere – hydrophobic interaction).

**Discussion.** An important decision to make when using PyRod is the MD simulation length. To estimate the equilibration process of water molecules in protein binding pockets, the change of water occupancy in trajectory bins of 50 frames was analyzed for each test system and plotted together with the RMSD of protein heavy atoms (supporting information Fig. S1). All test systems in this study were equilibrated within first 5 ns of unrestrained MD simulation. However, this may be different for other systems. Interestingly, these plots indicate a synchronous equilibration of protein and water rendering the protein RMSD an easy-to-use descriptor to estimate equilibration times of water molecules in protein binding pockets. The total simulation length was restricted to 10 ns for each replication to reduce computational costs and to sample protein conformations close to the crystallographic structure. Replications were performed to expand sampling of protein conformations without introducing artifacts from a single long MD simulation stuck in a local minimum[43].

Best PyRod pharmacophores of CDK2, D3R and HIV1P outperform the docking program DOCK 3.6 when comparing early enrichment factors (EF1) with the DUD-E benchmark[28] (supporting information Tab. S4). However, PyRod pharmacophores could not be generated for $A_{2A}R$ and ERα, since both targets quickly leave the ligand-bound conformation upon unrestrained MD simulation. Although restraining the protein heavy atoms is tempting, this procedure would neglect the contribution of the protein to the entropy of the system[44]. When restraining heavy atoms of the CDK2 system we observed many more stable hydration sites with overall higher feature scores, which may hinder prioritization of important chemical features (Figure 6). Instead of restraining the protein in MD simulations, it might be an option in such situation to employ methods that generate pharmacophores based on the static structure[45,46].

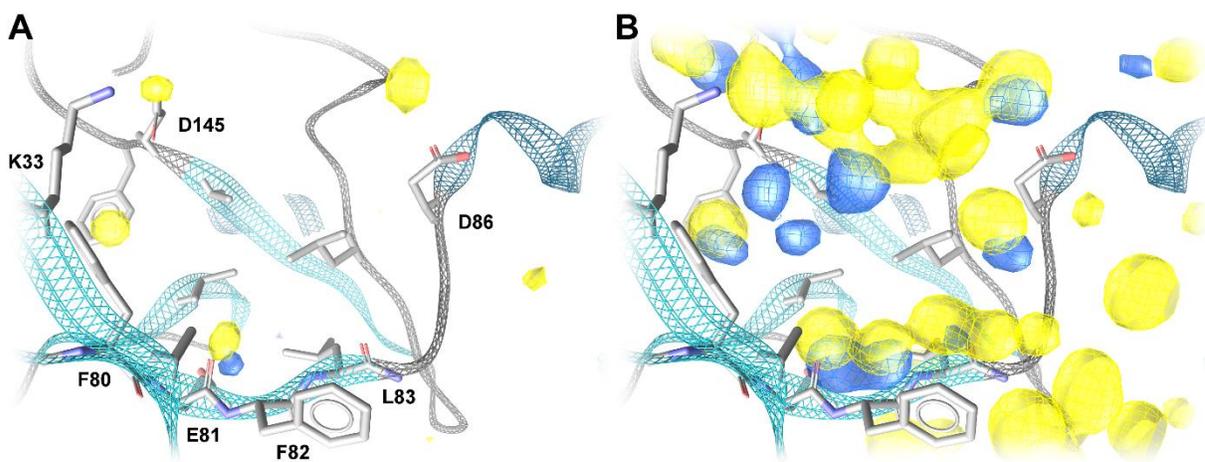

**Figure 6:** Effect of restraining protein heavy atoms on the binding pocket hydration sites of the CDK2 system. Generated dMIFs for hydrogen bonding interactions (HB) differ significantly between simulations with restrains on protein heavy atoms (yellow) and without (blue). HB

dMIFs show areas with a HB score of at least 90 % of the maximal HB score (A) and at least 50 % of the maximal HB score (B).

An important difference between the already published water pharmacophore method[8] and our PyRod approach is the number of generated pharmacophores. The water pharmacophore method was designed to generate a single pharmacophore in a highly automated fashion. Although retrospectively successful with 4 out of 7 targets, it needs to be shown that the parameters and cutoffs trained on the test systems also succeed in a prospective study on a completely different target. PyRod does not generate a single pharmacophore for virtual screening, but a combinatorial library. This agrees with the fact that different ligands can show different interaction patters for the same binding pocket. However, it would be desirable to develop only few diverse pharmacophore models with PyRod without knowledge of any ligand data. Prospective studies are on the way and the only possibility to proof the usefulness of PyRod in such situation. Nevertheless, we are confident that we would have been able generate successful pharmacophores by only analyzing dMIFs and selecting features for CDK2 and HIV1P.

CONCLUSION

In this study we could show that water dynamics from MD simulations can be used to generate highly usable 3D pharmacophore models for virtual screening. Employing the free and open-source software PyRod we were able to successfully describe pharmacophoric binding pocket characteristics and generate pharmacophores for three pharmaceutically relevant drug targets. The early enrichment factors from the best performing models range from 7.7 for D3R to 54.6 for HIV1P. Additionally, we found that restraining protein heavy atoms dramatically affects the water dynamics in the binding pocket hindering hot spot identification for ligand binding in water-based methods.

## ASSOCIATED CONTENT

**Supporting Information**. Tables for feature definitions, scoring functions, combinatorial library parameters and early enrichment factor comparison with the DUD-E benchmark (PDF).

## AUTHOR INFORMATION

**Corresponding Author**

*e-mail: gerhard.wolber@fu-berlin.de

**Author Contributions**

The manuscript was written through contributions of all authors. DS designed the software. DS, SP and GW designed experiments. DS and SP performed and analyzed experiments. GW directed the studies. All authors have given approval to the final version of the manuscript.

## ACKNOWLEDGMENT

We would like to thank the Elsa-Neumann-Foundation for supporting DS.

## ABBREVIATIONS

3D-RISM, 3D reference interaction site model; $A_{2A}R$, adenosine $A_{2A}$ receptor; AI, aromatic interaction; CDK2, cyclin-dependent kinase 2; D3R, dopamine D3 receptor; dMIF, dynamic molecular interaction field; ERα, estrogen receptor alpha; GIST, grid inhomogeneous solvation theory; HA, single hydrogen bond acceptor; HA2, double hydrogen bond acceptor; HB, hydrogen bond; HD, single hydrogen bond donor; HD2, double hydrogen bond donor; HDA, mixed hydrogen bond donor/acceptor; HI, hydrophobic interaction; HIV1P, human immunodeficiency virus 1 protease; MD simulation, molecular dynamics simulation; NI, negative ionizable; PI, positive ionizable; TW, trapped water.

(1) 5if1

(2) 1nh0

(3) 1xpc

(4) 3pbl

(5) 5iu4